# Improving Memory Utilization in Convolutional Neural Network Accelerators

Petar Jokic, *Member, IEEE*, Stephane Emery, *Member, IEEE*, and Luca Benini, *Fellow, IEEE*

*Abstract*—While the accuracy of convolutional neural networks has achieved vast improvements by introducing larger and deeper network architectures, also the memory footprint for storing their parameters and activations has increased. This trend especially challenges power- and resource-limited accelerator designs, which are often restricted to store all network data in on-chip memory to avoid interfacing energy-hungry external memories. Maximizing the network size that fits on a given accelerator thus requires to maximize its memory utilization. While the traditionally used *ping-pong* buffering technique is mapping subsequent activation layers to disjunctive memory regions, we propose a mapping method that allows these regions to overlap and thus utilize the memory more efficiently. This work presents the mathematical model to compute the maximum activations memory overlap and thus the lower bound of on-chip memory needed to perform layer-by-layer processing of convolutional neural networks on memory-limited accelerators. Our experiments with various real-world object detector networks show that the proposed mapping technique can decrease the activations memory by up to 32.9%, reducing the overall memory for the entire network by up to 23.9% compared to traditional *ping-pong* buffering. For higher resolution de-noising networks, we achieve activation memory savings of 48.8%. Additionally, we implement a face detector network on an FPGA-based camera to validate these memory savings on a complete end-to-end system.

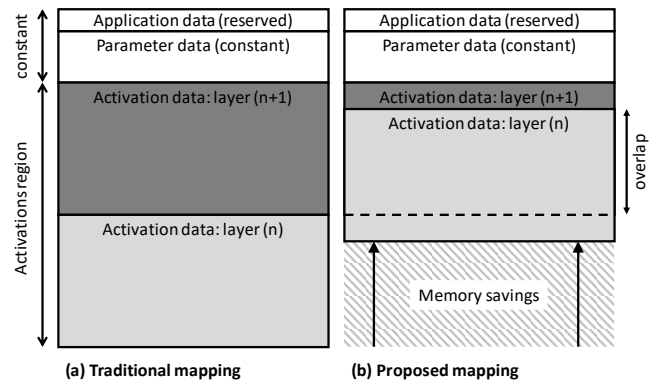

Fig. 1. Memory allocation of the traditional (a) and the proposed (b) activations mapping approach, visualizing the introduced overlap.

## I. INTRODUCTION

Convolutional neural networks (CNN) are the key components of today's state of the art object detectors and classifiers in computer vision. Performing inference with a CNN is a highly data-intensive task in which input activations, starting with an input image, get convolved with kernels consisting of learned weights, summed up with bias parameters and fed through an activation function. The layered structure of CNNs allows them to be processed sequentially, layer by layer. This is beneficial in terms of memory requirements because a maximum of only two subsequent activation layers, as opposed to all of them, have to be stored at any point in time: the inputs from the preceding layer are needed to be convolved with the kernels, while the results of these computations (output activations) must be buffered to serve as inputs for processing the following layer. Because consecutive layers output their results alternatingly into one of two activation memory sections this pattern is called *ping-pong* processing. The network parameters (weights and biases) are reused at every inference of the network and should thus be kept in local memory to avoid costly data reloading from external memory. To succeed in storing all network data on-chip for layer-wise CNN processing, the memory must be large enough to store the constant parameters and the largest pair of successive input and output activations as shown in Fig. 1 (a). With the traditionally used *ping-pong* buffering technique these activations are alternatingly mapped to disjunctive memory regions, such that the worst-case pair of activations amounts to the maximum sum of any two subsequent layers [1]. For CNN accelerators on resource-limited platforms, like field-programmable gate arrays (FPGA) [2, 3], this constraint largely limits the maximum network size that can be processed. On-chip static random-access memories (SRAM) dominate today's CNN accelerator designs (e.g. around 1 mm$^2$ for 1 MB in 22 nm technology) [4]. Reducing memory size will therefore clearly reduce chip area and thus largely influence the chip cost. Additionally, large memories increase the static power consumption due to leakage, and also their energy-per-access is heavily impacted by size, surpassing the energy consumed by the processing of the fetched data by a factor of more than 25x [5, 6]. Thus, it is essential to minimize the on-chip memory size to the targeted networks' needs.

This work presents a CNN memory mapping method that allows activation regions of subsequent layers to overlap, as shown in Fig. 1 (b), and thus utilize the memory more efficiently than the traditionally used *ping-pong* buffering technique. It consists of a mathematical model for computing the maximum overlap and thus its lower bound of on-chip memory needed to perform layer-wise processing of convolutional neural networks. This is especially attractive for newer networks where the memory is dominated by activations. The resulting memory size can be used to determine the minimum memory requirements for a new accelerator design or to optimize a given network to efficiently utilize the memory resources of an existing accelerator. Our experiments show activation memory savings of up to 32.9% for real-world object detector CNNs and up to 48.8% for high-resolution de-noising CNNs when compared to traditional ping-pong buffering.

## II. IMPROVING CNN MEMORY UTILIZATION

The traditional *ping-pong* mapping of activations ensures that the outputs of a layer do not overwrite any of its input activations, because they might still be needed for pending computations. But allocating two separate regions for this reason is too pessimistic, unnecessarily restricting the allowed network size, keeping the memory utilization low and thus the power consumption as well as cost high. In the following sections, we show how the data access pattern of CNNs can be exploited to improve the memory utilization of accelerators.

### A. CNN data access pattern

To determine the memory requirements for computing an entire CNN inference, we need to understand the data access pattern of this process. Fig. 2 visualizes the computational structure of a CNN layer, convolving an input feature map *in* (of size $X_{in} \cdot Y_{in} \cdot C_{in}$) with a weight kernel $k$ ($C_{out}$ kernels of size $K_x \cdot K_y \cdot C_{in}$), producing an output feature map *out* (of size $X_{out} \cdot Y_{out} \cdot C_{out}$). To convolve the whole input feature map, input activations are accessed in a sliding window operation, moving the kernel-sized window across the x/y plane. This operation can be represented with the 6 nested loops shown in Fig. 3. The window moves in strides of $s_x$ and $s_y$ in x- and y-direction, respectively. Inputs can be padded with $P_x$ and $P_y$ zero-pixels on each side in x- and y-direction.

Input data are stored in memory in the depth-first order: all $C_{in}$ input channels of an input pixel are followed by all entries of the neighboring input pixels in the x-direction. At the end of a row, the following rows in the y-direction are appended. This simplifies the addressing scheme and keeps the number of cycles between data reuse low by following the window pattern. Minimizing this so-called reuse distance [7] is important as it allows a specific memory entry to be overwritten as soon as possible, freeing space for new data. Because output feature maps will serve as inputs in the next layer, they must have the same memory order as the input feature maps.

### B. Model for optimized memory utilization

To optimize the memory utilization in CNN processors we propose a memory mapping method that allows activation memory regions of consecutive layers to be overlapping. Fig. 1 shows a simplified memory map that compares the *traditional* memory allocation (a) with our proposed approach (b). While (a) is keeping each layer's activations in separate disjunctive regions, (b) allows the activation regions to be partially overlapping, resulting in large memory savings.

If two subsequent layers have overlapping memory regions, the allocation method must avoid that output activations are overwriting data from the preceding layer that is still needed for pending computations. This constraint can be ensured by (negatively) offsetting the output write pointer in such a way, that the input reading pointer will never be reached during the computation of any layer in the network. This concept is depicted in Fig. 4. At the beginning of each layer computation (here denoted as t=0), the distance between the input activations read pointer $p_r$ and the output activations write pointer $p_w$, is set to an optimized offset. As the sliding window for computing the convolution operation moves on, pointer $p_w$ writes results and increments in direction of $p_r$. Pointer $p_r$ moves accordingly

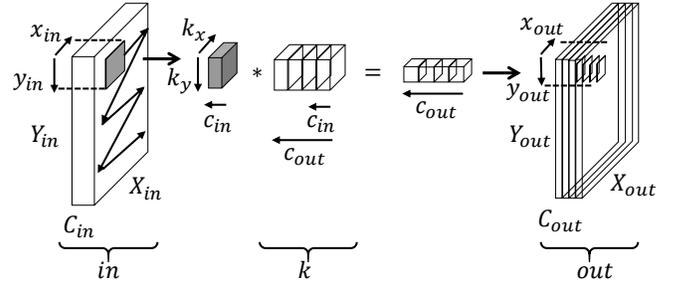

Fig. 2. Visualization of the 2D convolution operation in a CNN layer with relevant dimensions of the input and output activations as well as the kernel.

for $y_{out}$ in 0 to $Y_{out}(i)$
  for $x_{out}$ in 0 to $X_{out}(i)$
    for $c_{out}$ in 0 to $C_{out}(i)$
      for $k_y$ in 0 to $K_y(i)$
        for $k_x$ in 0 to $K_x(i)$
          for $c_{in}$ in 0 to $C_{in}(i)$
            $y_{in} = y_{out} \cdot s_y(i) - P_y(i)$
            $x_{in} = x_{out} \cdot s_x(i) - P_x(i)$
            out$(y_{out}, x_{out}, c_{out})$ += \
              in$(y_{in} + k_y, x_{in} + k_x, c_{in}) \cdot$ k$(k_y, k_x, c_{in}, c_{out})$

Fig. 3. Computation loops of a CNN layer i (omitting the accumulation reset, the input padding and the activation function at the end of each $c_{out}$ loop).

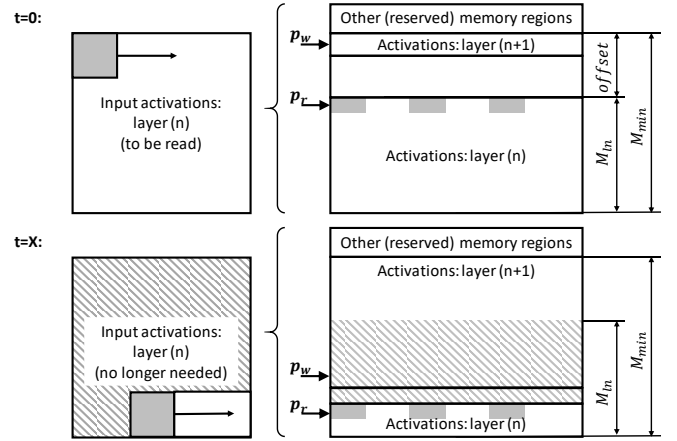

Fig. 4. Simplified memory map with current pointer positions at two different points in time. The left figures show the current position of the sliding window while the right figures visualize the memory content (including window data).

on the input activations region, away from $p_w$. The underlying idea of this memory mapping is the locality of the convolution operation: the activations for each window position are only read from a small, connected region of the input layer which itself is slid over the inputs in a continuous fashion. Because the corresponding memory data is ordered in the same way as they appear in this sliding operation, most parts of the processed input data will never be used again and can thus be overwritten by resulting output activations.

The maximum overlap of the two activation regions is found by mathematically describing the pointer positions and optimizing their relative offset distance at the beginning of each layer such that the total memory is minimal while constraining the write pointer to be smaller than the read pointer, avoiding any overwriting of still needed data. To meet this constraint, both pointer positions must be known for every point in time. They can be calculated from their starting points and velocities, derived from the network characteristics. We model the pointer positions (addresses) as a function of time, assuming one multiply-accumulate (MAC) operation per clock cycle (t) and that activation data is stored in the depth-first order described above. The sliding window follows this pattern and only moves on once all outputs for a certain window position are computed.

Equation (1) represents the velocity of pointer $p_w$, which advances 1 position per calculation of a single kernel convolution (or $C_{in} \cdot K_x \cdot K_y$ MAC operations). The address only increments once the full kernel is computed, which can be mathematically represented by rounding down the integral of its speed over time as shown in (2). The formula for $p_r$ takes the padding of the input layer, stride width, and the behavior during sliding window movements into account. It is sufficient to look at the lowest address of the input window (which would collide with $p_w$ at the earliest), simplifying the formula of its average velocity to (3). In the resulting $p_r$ formula (4), the y-direction stride is implemented with rounding operations, causing the pointer to skip some rows when moving in the y-direction. The minimum memory $M_{min,l_n}$ required for mapping the activations of two subsequent CNN layers to the memory can then be determined from (2) and (4), as shown in (5). It is given by the sum of the input activations space $M_{l_n}$ and the minimum offset difference $(p_{r0} - p_{w0})$ for which $p_r(t)$ is larger than $p_w(t)$ throughout the entire computation of a layer $l_n$ (during the interval $T_{l_n}$). From (5), the lower bound of activations memory required for computing the entire CNN, $M_{min}$, can be derived by finding the minimum memory size that supports all layers of the network, as shown in (6).

The worst-case scenario for memories in layer-wise CNN accelerators is a network with two maximum-sized layers back-to-back, requiring an activations buffer of twice the maximum layer size when using the traditional ping-pong buffering. For the same scenario, our method can reduce the memory needs by almost 50% if each input pixel creates equally many output pixels, keeping pointers at a constant short offset. This represents the theoretical upper savings limit of the proposed technique. Our model assumes one datum per memory word but can be easily transferred to multiple data entries per word by linearly scaling down the speed of each pointer accordingly. Note that for residual layers, $M_{l_n}$ must additionally include activations of identity connections in parallel to the convolutions.

### III. Experiments and Results

We evaluate the memory savings of the presented method on four real-world CNN networks: 9-layer DLIB face detector [8], 12-layer YOLO lite [9], 20-layer DMCNN-VD 3x3 [10] and 12-layer MobileNetv2 [11]. The first three have an input resolution of 640x640, while the input of MobileNetv2 is 224x224x3. Table I presents the memory savings of our proposed method compared to traditional *ping-pong* buffering. Our approach is saving between 19.6% and 48.8% of activations memory in the evaluated networks, achieving total memory savings (including parameters) of 6.2% to 48.2%. The lowest overall savings are found in MobileNetv2, where parameters dominate the memory due to the deep architecture and the small image size. It must be noted that we compute MobileNetv2 in a strictly layer-wise manner, while [11] suggests that operations of some intermediate layers could be concatenated without buffering the respective layers entirely. Many recent networks feature small kernels and larger images, increasing the dominance of activations in memory and thus memory savings. This can be seen in the 20-layer DMCNN-VD

$$v_{p_w} = \frac{1}{C_{in} \cdot K_x \cdot K_y} \tag{1}$$

$$p_w(t) = \text{floor}(v_{p_w} \cdot t) + p_{w0} \tag{2}$$

$$v_{p_r} = \underbrace{\frac{s_x \cdot C_{in}}{C_{out} \cdot (C_{in} \cdot K_x \cdot K_y)}}_{x-\text{speed}} + \underbrace{\frac{(s_y - 1) \cdot C_{in} \cdot X_{in}}{\left(\text{floor}\left(\frac{2 \cdot P_x + X_{in} - K_x}{s_x}\right) + 1\right) \cdot C_{out} \cdot (C_{in} \cdot K_x \cdot K_y)}}_{y-\text{speed}} \tag{3}$$

$$p_r(t) = \max\Bigg(0, \underbrace{\text{floor}\left(\frac{1}{C_{out} \cdot (C_{in} \cdot K_x \cdot K_y)} \cdot t\right) \cdot (s_x \cdot C_{in})}_{x \text{ position}} + \underbrace{\text{floor}\left(\frac{1}{\left(floor\left(\frac{2 \cdot P_x + X_{in} - K_x}{s_x}\right) + 1\right) \cdot C_{out} \cdot (C_{in} \cdot K_x \cdot K_y)} \cdot t\right) \cdot \left((s_y - 1) \cdot C_{in} \cdot X_{in}\right)}_{y \text{ position}} - \backslash$$

$$\underbrace{C_{in} \cdot X_{in} \cdot p_y}_{\text{top padding}} - \underbrace{\text{ceil}\left(\frac{1}{\left(floor\left(\frac{2 \cdot P_x + X_{in} - K_x}{s_x}\right) + 1\right) \cdot C_{out} \cdot (C_{in} \cdot K_x \cdot K_y)} \cdot t\right) \cdot \max\left(0, \left(\left(\text{floor}\left(\frac{2 \cdot P_x + X_{in} - K_x}{s_x}\right) + 1\right) \cdot s_x - X_{in}\right) \cdot (s_x \cdot C_{in})\right)}_{\text{side padding offset}}\Bigg) + p_{r0} \tag{4}$$

$$M_{min,l_n} = \min_{\substack{t \in T_{l_n} \\ p_r(t) > p_w(t)}} \left(M_{l_n} + (p_{r0} - p_{w0})\right) \quad T_{l_n} = \left[0, \left(\text{floor}\left(\frac{2 \cdot P_x + X_{in} - K_x}{s_x}\right) + 1\right) \cdot \left(\text{floor}\left(\frac{2 \cdot P_y + Y_{in} - K_y}{s_y}\right) + 1\right) \cdot C_{out} - 1\right] \tag{5}$$

$$M_{min} = \max_{l_n \in \mathcal{L}}(M_{min,l_n}) \tag{6}$$

with small 3x3 kernels, yielding 48.2% total memory savings. We note that our technique still offers significant (32.9%) activation memory savings for smaller networks, such as DLIB.

To validate the memory savings in a real application, we employ our method for implementing a DLIB face detector CNN [8] on a configurable FastEye camera [12]. This hosts a 1-megapixel image sensor and a Xilinx XC7K325T FPGA. We extend the existing data-path, implementing the sensor readout and a USB interface, with a simple CNN processing state machine and a block memory (BRAM) consisting of 36 kbit blocks for parameters and activations. The image from the sensor gets cropped to 640x640 pixels and stored on the BRAM. Triggered by the image, the state machine processes the network layer-wise as described in Fig. 3. Without any processing optimizations, the maximum CNN inference rate is 0.5 frames per second at 100 MHz clock frequency. Weights and activations are quantized to 16 bits. The output of the last CNN layer gets transferred via USB to a computer for post-processing of the resulting bounding boxes. Two different system configurations are implemented, a) with the standard *ping-pong* mapping, and b) with our proposed memory mapping technique, differing only in the BRAM size and the address generation. Both FPGA implementations successfully perform on-camera face detection on acquired images. Table II states the utilization report of the initial FPGA firmware and the two CNN-extended versions. Comparing the resources added to the initial camera firmware, the proposed memory mapping (b) shows memory savings of 23% and power savings of 20% with respect to the standard memory allocation (a). The number of used flip-flops (FF), look-up tables (LUT) and signal processors (DSP) in the FPGA rests almost constant. This confirms the theoretical savings (23.9%), differing by only 0.9%, which is due to the limited block granularity of the memory macro.

TABLE I
RESULTS OF EVALUATED NETWORKS

| Network name | CNN network | | | Mem. savings: activations only (total network) |
|---|---|---|---|---|
| | Parameter [# words] | Activations [# words] | | |
| | | Standard | This work | |
| DLIB face det. [8] | 229.8k | 614.4k | 412.2k | 32.9% (23.9%) |
| YOLO Lite [9] | 443.0k | 16.4M | 13.1M | 19.9% (18.7%) |
| MobileNetv2 [11] | 3.3M | 1.5M | 1.2M | 19.6% (6.2%) |
| DMCNN-VD [10] | 668.2k | 53.7M | 27.5M | 48.8% (48.2%) |

TABLE II
FPGA UTILIZATION REPORT AND POWER MEASUREMENT OF THE CAMERA

| | LUT | FF | DSP | BRAM | Power |
|---|---|---|---|---|---|
| **Cam only** | 28.6k (14%) | 82.4k (20%) | 8 (1%) | 12 (3%) | 12.61 W |
| **a) Cam + CNN** | 52.9k (26%) | 98.3k (24%) | 13 (2%) | 428 (96%) | 13.20 W |
| **b) Cam + CNN** | 52.5k (26%) | 98.3k (24%) | 13 (2%) | 332 (75%) | 13.08 W |
| **Savings: b vs. a** | 2% | 0% | 0% | 23% | 20 % |

## IV. RELATED WORK

Stoutchinin et al. [7] present an optimal model search approach that outputs optimized CNN loop order, tiling and buffer size parameters to reduce access to external memories. They achieve memory bandwidth reductions of up to 14x compared to previous implementations. Yang et al. [6] propose an analytical approach to model data locality in CNNs to find the optimal blocking strategy that maximizes the energy efficiency of an accelerator. Other works like [1] focus on optimizing data movements between internal and external memory while using traditional *ping-pong* buffering for on-chip memory. These approaches either do not consider cases with all activations stored on-chip or base their models on the inefficient *ping-pong* buffering. In contrast, we provide a more efficient activations mapping that can be used on any platform and only requires the adaption of the addressing scheme. While we focus on standard convolutions, networks with separable convolutions [11] allow intermediate layers to be stored only partially, reducing the memory footprint of intermediate layers with many channels.

## V. CONCLUSION

This work presented the mathematical model of the lower memory bound for buffering activations in layer-wise convolutional neural network accelerators using overlapping activation regions. We show that the mapping method derived from this model can utilize the memory more efficiently than the standard *ping-pong* buffering method. This allows reducing the required on-chip memory size of new accelerator designs or to map larger networks to existing resource-limited implementations. Experimental results on real-world CNNs show that the activations memory space can be reduced by up to 48.8%, and the overall network memory needs by up to 48.2%.